\documentclass[twocolumn,aps,superscriptaddress,nofootinbib,groupedaddress]{revtex4-1}

\usepackage{amsmath,amsthm,amsfonts}
\usepackage{graphicx}
\usepackage{physics}
\usepackage{hyperref}
\hypersetup{
	colorlinks   = true, %Colours links instead of ugly boxes
	urlcolor     = blue, %Colour for external hyperlinks
	linkcolor    = blue, %Colour of internal links
	citecolor    = red   %Colour of citations
}

\graphicspath{{figs/}}

\newcommand{\nc}{\newcommand}

\nc{\calR}{{\cal{R}}}
\nc{\calP}{{\cal{P}}}
\nc{\cN}{ {\cal{N}} }

\nc{\Mpt}{M_{_{\rm Pl}}^2}

\usepackage{tikz}
\usetikzlibrary{arrows,shapes}
\usetikzlibrary{trees}
\usetikzlibrary{matrix,arrows} 				% For commutative diagram
\usetikzlibrary{positioning}				% For "above of=" commands
\usetikzlibrary{calc,through}				% For coordinates
\usetikzlibrary{decorations.pathreplacing}  % For curly braces
\usepackage{pgffor}							% For repeating patterns

\usetikzlibrary{decorations.pathmorphing}	% For Feynman Diagrams
\usetikzlibrary{decorations.markings}
\tikzset{
    vector/.style={decorate, decoration={snake}, draw},
	provector/.style={decorate, decoration={snake,amplitude=2.5pt}, draw},
	antivector/.style={decorate, decoration={snake,amplitude=-2.5pt}, draw},
    fermion/.style={draw=black, postaction={decorate},
        decoration={markings,mark=at position .55 with {\arrow[draw=black]{>}}}},
    fermionbar/.style={draw=black, postaction={decorate},
        decoration={markings,mark=at position .55 with {\arrow[draw=black]{<}}}},
    fermionnoarrow/.style={draw=black},
    gluon/.style={decorate, draw=black,
        decoration={coil,amplitude=4pt, segment length=5pt}},
    scalar/.style={dashed,draw=black, postaction={decorate},
        decoration={markings,mark=at position .55 with {\arrow[draw=black]{>}}}},
    scalarbar/.style={dashed,draw=black, postaction={decorate},
        decoration={markings,mark=at position .55 with {\arrow[draw=black]{<}}}},
    scalarnoarrow/.style={dashed,draw=black},
    electron/.style={draw=black, postaction={decorate},
        decoration={markings,mark=at position .55 with {\arrow[draw=black]{>}}}},
	bigvector/.style={decorate, decoration={snake,amplitude=4pt}, draw},
}

\tikzstyle{block} = [draw, rectangle,
    minimum height=3em, minimum width=6em]

\begin{document}
	
\title{Planar Black Holes in Holographic Axion Gravity:\\ Islands, Page Times, and Scrambling Times }

\author{Seyed Ali Hosseini Mansoori$^{1}$}
\email{shosseini@shahroodut.ac.ir}

\author{Orlando Luongo$^{2,3,4,5}$}
\email{orlando.luongo@unicam.it}

\author{Stefano Mancini$^{2,5}$}
\email{stefano.mancini@unicam.it}

\author{Mirmani Mirjalali$^{1}$}
\email{mirmanimirjalali4@gmail.com }

\author{Morteza Rafiee$^{1}$}
\email{m.rafiee@shahroodut.ac.ir}

\author{Alireza Tavanfar$^{6,7}$}
\email{alireza.tavanfar@research.fchampalimaud.org}

\affiliation{$^{1}$Faculty of Physics, Shahrood University of Technology, P.O. Box 3619995161 Shahrood, Iran}
\affiliation{$^{2}$Scuola di Scienze e Tecnologie, Universit\`a di Camerino, Via Madonna delle Carceri 9, 62032 Camerino, Italy}
\affiliation{$^{3}$Dipartimento di Matematica, Universit\`a di Pisa, Largo B. Pontecorvo 5, Pisa, 56127, Italy}
\affiliation{$^{4}$NNLOT, Al-Farabi Kazakh National University, Al-Farabi av. 71, 050040 Almaty, Kazakhstan}
\affiliation{$^{5}$Istituto Nazionale di Fisica Nucleare, Sezione di Perugia, Via Alessandro Pascoli 23c, 06123 Perugia, Italy}
\affiliation{$^{6}$Champalimaud Research, Champalimaud Center for the Unknown, 1400-038 Lisboa, Portugal}
\affiliation{$^{7}$Institute of Neuroscience, University of Oregon, Eugene, OR 97403, USA}
	
\begin{abstract}
%The present paper aims to
The present work investigates the entanglement entropies of the Hawking radiations, the Page times, and the scrambling times, for the eternal planar black holes in the holographic axion gravity. The solutions correspond to a new class of charged black holes, because the boundary diffeomorphism is broken due to the graviton mass induced by the axion fields in the bulk. The information theoretical aspects of these black hole solutions is determined upon applying the island rule for the entanglement entropy.
Like non-extremal charged black holes, the radiation entropy grows linearly in the no-island configurations, while is saturated at late times %to the ?
by asymptotic values set by %\emph{the double of}
the Bekenstein-Hawking entropy in % the presence of
the island configurations, with the boundary being located slightly outside the outer horizon. In particular, for the extremal %planar
black planes of %our model
holographic axion gravity, we find that: (a) the entanglement entropy of the Hawking radiation is ill-defined at the early times when the island is absent; (b) it tends to a distinctive constant at the late times;
(c) the late-time location of the island is indeed universal. Moreover, we investigate how the Page time is affected by the holographic massive gravity deformation. For neutral solutions at the small deformation parameter, and for charged solutions with almost-extremal deformation parameter, we find that the Page transition happens at earlier times.
\end{abstract}

\maketitle %\pacs{98.80.Cq}

\section{Introduction}\label{sec0}
%%%%%% EM %%%%%%%%
Quantum information theoretic aspects and behaviour of gravitational horizons have served us as a long-lasting source of theoretically profound open questions. These questions emerge unavoidably in the overlapping physics of quantum field theory and Einstein's field theory of the curved spacetime. Typically, the standard semi-classical calculations lead to uncomfortable results which violate a principle fundamental to the known quantum physics: the time-evolution unitarity of closed quantum systems.
Hawking's initial semi-classical calculation suggests  an ongoing non-unitary process of black hole evaporation. %the entanglement (fine-grained or von-Neumann) entropy of Hawking radiation grows up to infinity during the  evaporation process, contrasting  with
The improved prescription by Page realizes a unitarity restoration with entropy decreasing right after the Page time \cite {Page:1993wv,Page:1976df, Page:2013dx}. Indeed, reproducing the Page curve for the entanglement entropy of Hawking radiation remains a key ingredient in solving conclusively the information problem in the gravitational quantum field theories.\\

Recently, a considerable insight has been gained within the ``\textit{island paradigm}" \cite{Engelhardt:2014gca,Penington:2019npb,
Almheiri:2019psf,
Almheiri:2019hni,
Almheiri:2019yqk,
Almheiri:2020cfm}. The entanglement entropy of the Hawking radiation, as one computes accordingly, may be limited during black hole evaporation \cite{Penington:2019npb,
Almheiri:2019psf,
Almheiri:2019hni,
Almheiri:2019yqk,
Almheiri:2020cfm}, realizing the Page curve and the time-evolution unitarity of all the process.
The fine-grained entanglement entropy of the Hawking radiation, in the spirit of quantum extremal surface prescription \cite{Engelhardt:2014gca,Almheiri:2019hni} and after renormalization  \cite{Bombelli:1986rw,Srednicki:1993im}, is given by,
 \begin{eqnarray}\label{GEformula2}
\quad S(\mathcal{R})
 =
 \rm min \left\{\mathop{\mathrm{ext}}_{\mathcal{I}}\left[
S_{\rm gen}
 \right]\right\}.&&
\end{eqnarray}
It is defined in terms of the generalized entropy $S_{\rm gen}  = \mathrm{Area}(\partial {\mathcal{I}})/4G_N
 + S_{\rm m}(\mathcal{R} \cup {\mathcal{I}})$
with $\mathcal{I}$ and $\mathcal{R}$ denoting in order the island and radiation regions, and $G_{N}$ being the Newton constant. Additionally, $\partial \mathcal{I}$ represents the boundary of the island, the quantum extremal surface, and the matter entropy $S_{\rm m}$
 is the von Neumann entropy of the quantum fields living on the union of the island and the radiation regions. This formula reveals that one first needs to extremize the generalized entropy to find extremal points which indicate locations of the island. Then one selects the minimum value as the fine-grained entropy of radiation. Interestingly, Eq. \eqref{GEformula2} can be also derived by means of the replica method applied to the gravitational path integral \cite{Penington:2019kki,Almheiri:2019qdq}. Moreover, the island formula can be understood by combining the AdS/BCFT correspondence and the brane world holography \cite{Almheiri:2019psy,Geng:2020qvw,Sully:2020pza,Chen:2020uac,Chen:2020hmv,Hernandez:2020nem,Grimaldi:2022suv,Suzuki:2022xwv, Bousso:2021sji,Bhattacharya:2021jrn,Bhattacharya:2021dnd,Bhattacharya:2021nqj,Hu:2022ymx,Hu:2022zgy,Geng:2020fxl,Geng:2021iyq}.\\

Preliminary works in this field  focused on reproducing successfully Page curve in two-dimensional black holes
 using the semiclassical method in Jackiw-Teitelboim (JT) gravity \cite{Almheiri:2019hni,Almheiri:2019qdq,Almheiri:2019yqk}. The island rule was generalized to the entanglement entropy of the Hawking radiation in higher dimensional black holes. For instance, the Schwarzschild \cite{Hashimoto:2020cas,Arefeva:2021kfx}, Reissner-Nordstr$\rm \ddot{o}$m \cite{Wang:2021woy}, Charged and neutral (generalized) dilaton \cite{Yu:2021cgi,Ahn:2021chg,Anegawa:2020ezn},  Kaluza-Klein \cite{Lu:2021gmv}, rotating BTZ \cite{Yu:2021rfg} black holes, Kerr-de Sitter spacetime \cite{Azarnia:2022kmp}, the black holes in massive gravity \cite{Cao:2021ujs} and in the presence of higher derivative gravity terms \cite{Alishahiha:2020qza,Yadav:2022fmo}, and hyperscaling violating black branes \cite{Omidi:2021opl}. For further developments in this direction, see also the non-exhaustive list \cite{Matsuo:2020ypv,Wang:2021mqq,Bak:2020enw,Geng:2020qvw,Ling:2020laa,Kim:2021gzd} and the references therein. \\

An increasing number of studies have been carried out on the island proposal in the higher dimensional de Sitter space \cite{Geng:2021wcq} and  cosmological scenarios \cite{Sybesma:2020fxg,Balasubramanian:2020xqf,Hartman:2020khs,Espindola:2022fqb,Aalsma:2021bit,Bousso:2022gth,Marolf:2021ghr,Azarnia:2021uch} and with other quantum information measures such as the entanglement negativity, mutual information, and relative entropy \cite{Shapourian:2020mkc,Kudler-Flam:2021rpr,Saha:2021ohr,RoyChowdhury:2022awr,KumarBasak:2021rrx,Afrasiar:2022ebi,KumarBasak:2020ams,Basu:2022reu}. Despite remarkable achievements of the island paradigm, notable counterexamples \cite{Li:2021lfo,Kames-King:2021etp} were shown against the general ability of the island formula to recover the Page curve and solve the information problem. However, in \cite{Tian:2022pso}, the author found a proper island in Liouville black hole which could solve the puzzle appeared in Ref. \cite{Li:2021lfo}. Having these questions in regard, it is interesting to examine the uncovered aspects of the Island perspective for other black hole geometries. \\

 In this work, in the context of the holographic massive gravity  \cite{Baggioli:2014roa,Alberte:2015isw} and by applying the island rule, we attempt to compute the entanglement entropy of the Hawking radiation and the Page curve for the AdS black holes which are coupled to two auxiliary flat baths. It should be noted that the coupling to such a non-gravitational baths can induce a mass to the bulk graviton \cite{Geng:2020qvw,Porrati:2001db} so that
the graviton mass is essential for the existence of island \cite{Geng:2020fxl,Geng:2021hlu}. However, here we only consider the graviton mass arising from the holographic massive gravity setup and study its effect on the Page curve. In fact, in the holographic massive gravity models, giving a mass to the graviton in the bulk, results in the momentum no longer being conserved at the boundary. Strictly speaking, the bulk graviton mass breaks the diffeomorphism invariance of the gravitational action and thus, from holographic point of view, the stress-energy tensor of the dual field theory is not conserved and in turn the momentum conservation will be violated (For a nice  comprehensive review on the holographic axion models and their applications, see Refs. \cite{Baggioli:2021xuv,Baggioli:2019rrs}.). A striking feature of this model is that the black plane solutions of this model, even in the absence of Maxwell tensor, have an inner horizon due to the finite graviton mass. Namely they play the same rule like charged black holes with two horizons. Therefore, inspired by  \cite{Wang:2021woy,Ahn:2021chg,Lu:2021gmv,Cao:2021ujs,Kim:2021gzd}, we can perform the separate island analysis for the extremal and the non-extremal cases. \\

This paper consists of five sections. Section \ref{sec2} begins with investigating neutral, non-extremal, and extremal planar black holes in the framework of the holographic massive gravity. In Sections \ref{sec3} and \ref{sec4}, we will obtain the entanglement entropy of the Hawking radiation for all kind of the planar black holes with and without the island. In addition, the scrambling time and the Page times are both analyzed and discussed, specially at the end of Section \ref{sec3}. Finally, our conclusions are drawn in Section \ref{sec5}. Thermodynamics of the planar black holes are detailed in Appendix \ref{sec: sec6}.

%%%%%%%%%%%%%%%%%%%%%%%%%%%%%%%%%%%%%%%%%%%%%%%%%%%%%
  \section{ Black plane solutions} \label{sec2}
  The aim of this section is to obtain a general family of non-extremal, extremal, and also neutral black plane solutions in holographic massive gravity.
 The action we consider for $(1+3)$-dimensional spacetimes is formulated as follows, \cite{Baggioli:2014roa,Alberte:2015isw,Baggioli:2021xuv,Alberte:2017oqx,Gouteraux:2016wxj,Baggioli:2016oqk,Baggioli:2021ejg}
  \begin{equation}\label{action1}
  S=\frac{1}{16 \pi G_N} \int dx^4 \sqrt{-g} \Big[R+\frac{6}{L^2} -K(X)\Big],
\end{equation}
%$G$ is the Newton constant and
 In the above action, $R$ is the Ricci scalar, $L$ denotes the AdS radius, and $X=\frac{1}{2} \sum_{I=1}^{2} \partial_{\mu} \Phi^{I}\partial^{\mu} \Phi^{I} $ is the kinetic term for the axions  $\Phi^{I}$ ($I=x,y$) which are shift-invariant massless scalar fields. Notice that $K$ is a generic scalar function \cite{Baggioli:2014roa,Baggioli:2021xuv,Alberte:2017oqx}. Bulk axion fields are described by the linear solution $\Phi^{I}=\alpha \; x^{I}$, with $\alpha$ being a constant. Interestingly, these scalars break translational invariance at the boundary, as they induce an effective graviton mass.
By varying the action with respect to $g_{\mu \nu}$, one arrives at,
\begin{eqnarray}\label{Ein1}
 G_{\mu \nu}-3 g_{\mu \nu} = \frac{1}{2} \Big[K'(X)\sum_{I=1}^2 \partial_{\mu} \Phi^{I} \partial_{\nu} \Phi^{I}
 - K(X)g_{\mu \nu}\Big]
\end{eqnarray}
with $G_{\mu \nu}$ being the Einstein tensor. We have chosen to use the units in which both $L = 1$ and $8 \pi G_{N} = M_{p}^{-2}=1$. In order to find a general family of static black planes, we consider the following ansatz for the solutions,
\begin{equation}\label{metric12}
ds^2=-r^2 f(r)dt^2+\frac{dr^2}{r^2 f(r)}+r^2(dx^2+dy^2)
\end{equation}
The blackening function, $f(r)$, vanishes at the outer event horizon which is located at $r_{+}$. Substituting the above ansatz in Eq. \eqref{Ein1}, the metric elements take the following general form \cite{Baggioli:2014roa,Gouteraux:2016wxj,Baggioli:2016oqk},
\begin{equation}\label{metricelement}
f(r)=1-\Big(\frac{r_{+}}{r}\Big)^3-\frac{1}{2r^3} \int_{r_{+}}^{r} ds s^2 K\Big(\frac{\alpha^2}{s^2}\Big)
\end{equation}
The associated temperature, $T= \kappa_{+}/(2 \pi)$ where $\kappa_{+}$ is the surface gravity of the outer horizon, is accordingly,
\begin{equation}\label{tem}
T = \frac{(r^2 f(r))'}{4 \pi}|_{r=r_{+}}=\frac{r_{+}}{8 \pi } \Big[6-K\Big(\frac{\alpha^2}{r_{+}^2}\Big)\Big]
\end{equation}
For neutral black planes corresponding to the massless graviton, $\alpha=0$, the temperature is $T=3 r_{+}/(4 \pi) $ \cite{Cai:1996eg}. We refer readers to Appendix \ref{sec: sec6} for more details about the thermodynamics of our black plane solution.\\\\
We are interested in black plane solutions for a power function, i.e. $K(X)=X^n$ with $n>0$, in order to avoid the
ghost and gradient instabilities \cite{Alberte:2017oqx}. In this case, Eq. \eqref{metricelement} yields,
%\begin{eqnarray}\label{frn}
%	f(r)=
%	\frac{1}{r^3}(r-r_+)Q(r) \nonumber
%\end{eqnarray}
%with,
%\begin{eqnarray}
%\nonumber Q(r)= \mathcal{I}_{2}(r) -  \frac{\alpha^{2n}}{6-4 n} \Bigg\{\begin{matrix}
%	 \mathcal{I}_{3-2n}(r) & \text{for} & n< \frac{3}{2}\\
%	-\frac{1}{(rr_{+})^{2n-3}}\mathcal{I}_{2n-3}(r) & \text{for} & n> \frac{3}{2}&&
%\end{matrix}
%\end{eqnarray}
%where $\mathcal{I}_{l}(r)=\sum_{m=1}^{l} r^{l-m} r_+^{m-1}$ for $l>0$
\begin{equation}\label{frn}\begin{split}
&	f(r)=
	\frac{1}{r^3}(r-r_+)Q(r)  \\
& Q(r)= \mathcal{N}_{2}(r) -  \frac{\alpha^{2n}}{6-4 n} \Bigg\{\begin{matrix}
	 \mathcal{N}_{3-2n}(r) & \text{for} & n< \frac{3}{2}\\
	-\frac{1}{(rr_{+})^{2n-3}}\mathcal{N}_{2n-3}(r) & \text{for} & n> \frac{3}{2}&&
\end{matrix}\\
& \mathcal{N}_{l}(r)=\sum_{m=1}^{l} r^{l-m} r_+^{m-1}\;\;\;;\;\; l \geq 1
\end{split} \end{equation}
The solution\footnote{In the case $n=3/2$ the metric function in Eq. \eqref{metricelement} takes the logarithmic form as
\begin{equation}
f(r)=1-\Big(\frac{r}{r_{+}}\Big)^{3}-\frac{\alpha^3}{r^3} \ln\Big(\frac{r}{r_{+}}\Big)
\end{equation}
which indicates a black plane solution with only one horizon at $r=r_{+}$.}
\eqref{frn} implies that there is always one outer horizon at $r=r_{+}$ corresponding to $f(r_{+})=0$, whereas the existence of the inner (Cauchy) horizon, $Q(r_{-})=0$, depends directly on the existence of the real root of $Q$. In the case of a solution with two horizons,  the metric function can be expressed as, % in the form, as a function of $(r_{-},r_{+})$,

\begin{eqnarray}\label{frmrp}
\nonumber f(r)&=& \frac{1}{r^3} (r-r_{+})\Bigg[\mathcal{N}_{2}(r)\\
&-& \mathcal{N}_{2}(r_{-})\Bigg\{\begin{matrix}
 \frac{\mathcal{N}_{3-2n}(r)}{\mathcal{N}_{3-2n}(r_{-})}& \text{for} & n<\frac{3}{2} \\
\Big(\frac{r_{-}}{r}\Big)^{2n-3} \frac{\mathcal{N}_{2n-3}(r)}{\mathcal{N}_{2n-3}(r_{-})}& \text{for} & n>\frac{3}{2}
\end{matrix}\Bigg]
\end{eqnarray}
It is worth mentioning that since the line element of spacetime must be spacelike between the horizons, the above solution is true as long as the first derivative of %the metric element
$f(r)$ is positive at $r_{+}$ and negative at $r_{-}$, that is $f'(r_{+})>0$\footnote{One requires this condition for having positive temperatures for the system.}  and $f'(r_{-})<0$. Nevertheless, one can find a special range of the mass generating parameter $\alpha$ in which the inner horizon can be absent. To make it clear, let us check this point in cases with $K(X) \in \{X,X^2,X^{3}\}$.

\begin{itemize}
\item \textbf{Type I}: $K(X) = X$ model
\end{itemize}
The case $K(X)=X$ corresponds to the well-known holographic model of momentum relaxation \cite{Andrade:2013gsa}. It is easy to show that
 the inner horizon is located at $r_-=\frac{1}{2} \left(\sqrt{2 \alpha^2-3 r_+^2}-r_+\right)$, for $\alpha$ values in the range $\sqrt{2} r_+ < \alpha < \sqrt{6} r_+$ according to  $f'(r_{+})>0$. Therefore, the metric function \eqref{frmrp} can be expressed as,
   \begin{equation}\label{fn1}
     	f(r)=\frac{1}{r^3}(r - r_+) (r - r_-) (r + r_+ + r_-)
   \end{equation}
   It should be noted that outside the mentioned range for $\alpha$, this solution has only one horizon and the Cauchy horizon will be absent.
\begin{itemize}
\item \textbf{Type II}: $K(X) = X^2$ model
\end{itemize}
	For this case, in the range $0 < \alpha^4 < 6 r_+^4$, we have always a solution with two horizons which is given by,
\begin{eqnarray}
	\nonumber f(r)&=& \frac{(r - r_+) (r - r_-)}{r^4} \Big(r^2 + r_+^2
	+ r_+ r_- + r_-^2 \\
	&+& r (r_+ + r_-)\Big)
\end{eqnarray}
Note that the temperature  is positive in the mentioned range.
\begin{itemize}
\item \textbf{Type III}: $K(X) = X^3$ model
\end{itemize}
Finally for $K(X) =  X^{3}$, the black plane possesses two event horizons at $r=r_+$ and $r=r_-=\frac{\alpha^2}{6^{1/3} r_+}$ for %ever in
the range $0<\alpha^6< 6 r_{+}^6$. The metric can also be written as,
\begin{equation}
f(r)=\frac{1}{r^6}(r^3-r_{+}^3)(r^3-r_{-}^3)
\end{equation}
 Generally, the upper bounds of the $\alpha$ ranges come from the positiveness of the temperature. Using Eq. \eqref{tem}, in the $K(X)=X^{n}$ models, one can obtain,
 \begin{equation}\label{alphiex}
 \alpha_{ext}=6^{\frac{1}{2n}} r_{+}
\end{equation}
It is worth mentioning that for $\alpha$ values near this bound, the black planes have two horizons and exactly at the $\alpha_{ext}$, the temperature is zero which implies the \textit{extremal} limit.
Before leaving this section, we should consider the extremal limit of our cases. In this limit two horizons $r_{\pm}$ coincide together and hence the black hole possesses only one horizon at $r_{+}=r_{-}=r_{e}$.
In this respect, the metric function \eqref{frmrp} becomes,

\begin{eqnarray}\label{frmre}
\nonumber &f_{e}(r)&=\frac{1}{r^3} (r-r_{e})\Bigg[\mathcal{N}^{e}_{2}(r)\\
&-&\frac{3 r_{e}^{2n}}{3-2n}\Bigg\{\begin{matrix}
 \mathcal{N}_{3-2n}(r)& \text{for} & n<\frac{3}{2} \\
 -(r r_{e})^{3-2n}\mathcal{N}_{2n-3}(r)& \text{for} & n>\frac{3}{2}
\end{matrix}\Bigg]\,\,
\end{eqnarray}
where $\mathcal{N}_{l}^e(r)=\sum_{m=1}^{l} r^{l-m} r_e^{m-1}$. For an example, for the $K=X$ case if $\alpha=\sqrt{6} r_{+}$, the metric function of the extremal black plane \eqref{frmre} recasts the form,
  \begin{equation}\label{emetric}
  f_{e}(r)=\frac{1}{ r^3}(r-r_{e})^2( r+2 r_{e})
  \end{equation}

%%%%%%%%%%%%%%%
\section{Entanglement entropy}\label{sec3}
In this section, we attempt to derive the entanglement entropy of the Hawking radiation for our black branes with or without the island configuration. Based on such calculations, we can reproduce the Page curves discussed in the context of black hole information paradox, with the Page times and scrambling times for our case of study.
\subsection{Entropy without Island
}
\begin{figure}
\begin{tikzpicture}[line width=1.5 pt, scale=1.1]
	\draw[vector] (2/1.3,2/1.3)--(-2/1.3,2/1.3);
	\draw[vector] (2/1.3,-2/1.3)--(-2/1.3,-2/1.3);
	\draw[dashed] (2/1.3,-2/1.3)--(-2/1.3,2/1.3);
	\draw[dashed] (2/1.3,2/1.3)--(-2/1.3,-2/1.3);
	\draw[red] (4/1.3,0) -- (2/1.3,-2/1.3);
	\draw[red] (4/1.3,0) -- (2/1.3,2/1.3);
	\draw[red] (-4/1.3,0) -- (-2/1.3,-2/1.3);
	\draw[red] (-4/1.3,0) -- (-2/1.3,2/1.3);
	\draw[black] (-2/1.3,2/1.3) -- (-2/1.3,-2/1.3);
	\draw[black] (2/1.3,2/1.3) -- (2/1.3,-2/1.3);
\node at (2.5/1.3,1/1.3) {$b_{+}$};	
\draw[blue,fill= blue] (2.5/1.3,0.75/1.3) circle (.035cm);
\draw[green,dotted] (4/1.3, 0) .. controls (3/1.3, 0.2/1.3) .. (2.5/1.3,0.75/1.3 );

\node at (-2.5/1.3,1/1.3) {$b_{-}$};	
\draw[blue,fill= blue] (-2.5/1.3,0.75/1.3) circle (.035cm);

\draw[green,dotted] (-4/1.3, 0) .. controls (-3/1.3, 0.2/1.3) .. (-2.5/1.3,0.75/1.3 );

\node at (0,2.3/1.3) {\small $r=0$};
\node at (0,-2.3/1.3) {\small $r=0$};

\node[rotate=45] at (1.3/1.3,1/1.3) {\small $r=r_{+}$};	
\node[rotate=-45] at (-1.3/1.3,1/1.3) {\small $r=r_{+}$};

\node[rotate=90] at (-2.2/1.3,0) {\small $r=\infty$};	
\node[rotate=90] at (2.2/1.3,0) {\small $r=\infty$};

\node at (0,-2.6/1.2) {\small $\text{(a) \textbf{Without island}}$};

\node[green] at (3/1.3,-0.1/1.3) {$\mathcal{R}_{+}$};
\node[green] at (-3/1.3,-0.1/1.3) {$\mathcal{R}_{-}$};	
\begin{scope}[shift={(0,-4.5)}]
\draw[vector] (2/1.3,2/1.3)--(-2/1.3,2/1.3);
	\draw[vector] (2/1.3,-2/1.3)--(-2/1.3,-2/1.3);
	\draw[dashed] (2/1.3,-2/1.3)--(-2/1.3,2/1.3);
	\draw[dashed] (2/1.3,2/1.3)--(-2/1.3,-2/1.3);
	\draw[red] (4/1.3,0) -- (2/1.3,-2/1.3);
	\draw[red] (4/1.3,0) -- (2/1.3,2/1.3);
	\draw[red] (-4/1.3,0) -- (-2/1.3,-2/1.3);
	\draw[red] (-4/1.3,0) -- (-2/1.3,2/1.3);
	\draw[black] (-2/1.3,2/1.3) -- (-2/1.3,-2/1.3);
	\draw[black] (2/1.3,2/1.3) -- (2/1.3,-2/1.3);
\node at (2.5/1.3,1/1.3) {$b_{+}$};	
\draw[blue,fill= blue] (2.5/1.3,0.75/1.3) circle (.035cm);
\draw[green,dotted] (4/1.3, 0) .. controls (3/1.3, 0.2/1.3) .. (2.5/1.3,0.75/1.3 );

\node at (-2.5/1.3,1/1.3) {$b_{-}$};	
\draw[blue,fill= blue] (-2.5/1.3,0.75/1.3) circle (.035cm);

\draw[green,dotted] (-4/1.3, 0) .. controls (-3/1.3, 0.2/1.3) .. (-2.5/1.3,0.75/1.3 );

\node at (0,2.3/1.3) {\small $r=0$};
\node at (0,-2.3/1.3) {\small $r=0$};

\node[rotate=45] at (-1.3/1.3,-1/1.3) {\small $r=r_{+}$};	
\node[rotate=-45] at (1.3/1.3,-1/1.3) {\small $r=r_{+}$};

\node at (0,-2.6/1.2) {\small $\text{(b)\textbf{ With island}}$};	

\node at (1.5/1.3,1/1.3) {$a_{+}$};	
\draw[blue,fill= blue] (1.5/1.3,0.75/1.3) circle (.035cm);

\node at (-1.5/1.3,1/1.3) {$a_{-}$};	
\draw[blue,fill= blue] (-1.5/1.3,0.75/1.3) circle (.035cm);

\draw[orange] (1.5/1.3,0.75/1.3) .. controls (0, 0.2) .. (-1.5/1.3,0.75/1.3);

\node[rotate=90] at (-2.2/1.3,0) {\small $r=\infty$};	
\node[rotate=90] at (2.2/1.3,0) {\small $r=\infty$};

\node[green] at (3/1.3,-0.1/1.3) {$\mathcal{R}_{+}$};
\node[green] at (-3/1.3,-0.1/1.3) {$\mathcal{R}_{-}$};	
\end{scope}
\end{tikzpicture}
\caption{(a) The Penrose diagram of the non-extremal neutral black plane solutions without island in the presence of a flat thermal bath (red lines). The Hawking radiation-identified area $\mathcal R$ is divided into two portions, $\mathcal R_+$ and $\mathcal R_-$, which are located in the right and the left wedges, respectively. The $\mathcal R_+$ and $\mathcal R_-$ boundary surfaces are denoted respectively by $b_+$ and $b_-$. (b) The Penrose diagram of the black plane solutions with island in the presence of a flat thermal bath (red lines) \cite{Ling:2020laa}. The boundaries of the island are located at $a_{-}$ and $a_+$. \label{fig1}}
\end{figure}
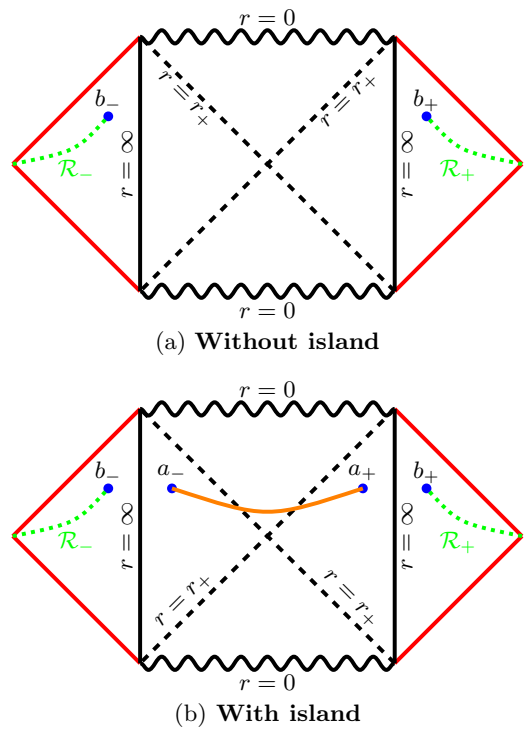

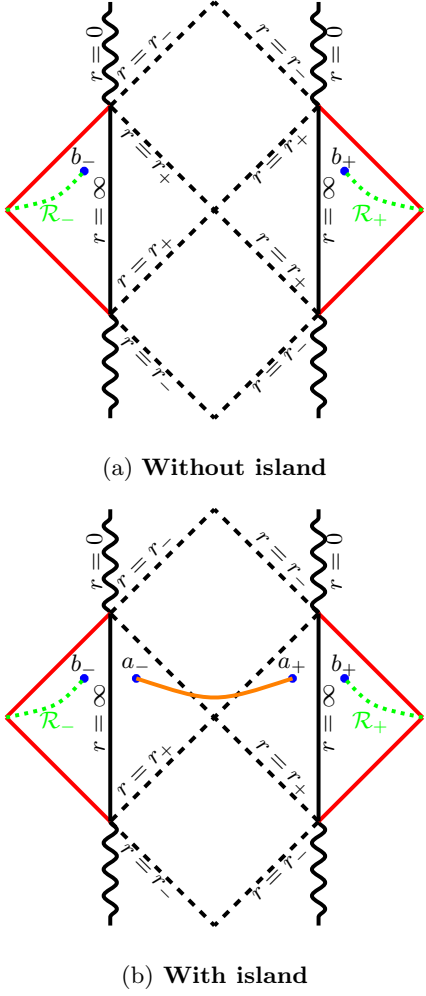
\begin{figure}
\begin{tikzpicture}[line width=1.5 pt, scale=0.9]
	\draw[vector] (2/1.3,2/1.3)--(2/1.3,4/1.3);
	\draw[vector] (-2/1.3,2/1.3)--(-2/1.3,4/1.3);
	\draw[vector] (2/1.3,-2/1.3)--(2/1.3,-4/1.3);
	\draw[vector] (-2/1.3,-2/1.3)--(-2/1.3,-4/1.3);
	\draw[dashed] (2/1.3,-2/1.3)--(-2/1.3,2/1.3);
	\draw[dashed] (2/1.3,2/1.3)--(0,4/1.3);
	\draw[dashed] (-2/1.3,2/1.3)--(0,4/1.3);
	\draw[dashed] (2/1.3,-2/1.3)--(0,-4/1.3);
	\draw[dashed] (-2/1.3,-2/1.3)--(0,-4/1.3);
	\draw[dashed] (2/1.3,2/1.3)--(-2/1.3,-2/1.3);
	\draw[red] (4/1.3,0) -- (2/1.3,-2/1.3);
	\draw[red] (4/1.3,0) -- (2/1.3,2/1.3);
	\draw[red] (-4/1.3,0) -- (-2/1.3,-2/1.3);
	\draw[red] (-4/1.3,0) -- (-2/1.3,2/1.3);
	\draw[black] (-2/1.3,2/1.3) -- (-2/1.3,-2/1.3);
	\draw[black] (2/1.3,2/1.3) -- (2/1.3,-2/1.3);
\node at (2.5/1.3,1/1.3) {$b_{+}$};	
\draw[blue,fill= blue] (2.5/1.3,0.75/1.3) circle (.035cm);
\draw[green,dotted] (4/1.3, 0) .. controls (3/1.3, 0.2/1.3) .. (2.5/1.3,0.75/1.3 );

\node at (-2.5/1.3,1/1.3) {$b_{-}$};	
\draw[blue,fill= blue] (-2.5/1.3,0.75/1.3) circle (.035cm);

\draw[green,dotted] (-4/1.3, 0) .. controls (-3/1.3, 0.2/1.3) .. (-2.5/1.3,0.75/1.3 );

\node[rotate=90] at (-2.3/1.3,3/1.3) {\small $r=0$};
\node[rotate=90] at (2.3/1.3,3/1.3) {\small $r=0$};

\node[rotate=45] at (1.3/1.3,1/1.3) {\small $r=r_{+}$};	
\node[rotate=-45] at (-1.3/1.3,1/1.3) {\small $r=r_{+}$};

\node[rotate=-45] at (1.3/1.3,-1/1.3) {\small $r=r_{+}$};	
\node[rotate=+45] at (-1.3/1.3,-1/1.3) {\small $r=r_{+}$};

\node[rotate=-45] at (1.3/1.3,3/1.3) {\small $r=r_{-}$};	
\node[rotate=45] at (-1.3/1.3,3/1.3) {\small $r=r_{-}$};

\node[rotate=45] at (1.3/1.3,-3/1.3) {\small $r=r_{-}$};	
\node[rotate=-45] at (-1.3/1.3,-3/1.3) {\small $r=r_{-}$};

\node[rotate=90] at (-2.2/1.3,0) {\small $r=\infty$};	
\node[rotate=90] at (2.2/1.3,0) {\small $r=\infty$};

\node at (0,-4.6/1.2) {\small $\text{(a) \textbf{Without island}}$};

\node[green] at (3/1.3,-0.1/1.3) {$\mathcal{R}_{+}$};
\node[green] at (-3/1.3,-0.1/1.3) {$\mathcal{R}_{-}$};		
\begin{scope}[shift={(0,-7.5)}]
\draw[vector] (2/1.3,2/1.3)--(2/1.3,4/1.3);
	\draw[vector] (-2/1.3,2/1.3)--(-2/1.3,4/1.3);
	\draw[vector] (2/1.3,-2/1.3)--(2/1.3,-4/1.3);
	\draw[vector] (-2/1.3,-2/1.3)--(-2/1.3,-4/1.3);
	\draw[dashed] (2/1.3,-2/1.3)--(-2/1.3,2/1.3);
	\draw[dashed] (2/1.3,2/1.3)--(0,4/1.3);
	\draw[dashed] (-2/1.3,2/1.3)--(0,4/1.3);
	\draw[dashed] (2/1.3,-2/1.3)--(0,-4/1.3);
	\draw[dashed] (-2/1.3,-2/1.3)--(0,-4/1.3);
	\draw[dashed] (2/1.3,2/1.3)--(-2/1.3,-2/1.3);
	\draw[red] (4/1.3,0) -- (2/1.3,-2/1.3);
	\draw[red] (4/1.3,0) -- (2/1.3,2/1.3);
	\draw[red] (-4/1.3,0) -- (-2/1.3,-2/1.3);
	\draw[red] (-4/1.3,0) -- (-2/1.3,2/1.3);
	\draw[black] (-2/1.3,2/1.3) -- (-2/1.3,-2/1.3);
	\draw[black] (2/1.3,2/1.3) -- (2/1.3,-2/1.3);
\node at (2.5/1.3,1/1.3) {$b_{+}$};	
\draw[blue,fill= blue] (2.5/1.3,0.75/1.3) circle (.035cm);
\draw[green,dotted] (4/1.3, 0) .. controls (3/1.3, 0.2/1.3) .. (2.5/1.3,0.75/1.3 );

\node at (-2.5/1.3,1/1.3) {$b_{-}$};	
\draw[blue,fill= blue] (-2.5/1.3,0.75/1.3) circle (.035cm);

\draw[green,dotted] (-4/1.3, 0) .. controls (-3/1.3, 0.2/1.3) .. (-2.5/1.3,0.75/1.3 );

\node[rotate=-45] at (1.3/1.3,3/1.3) {\small $r=r_{-}$};	
\node[rotate=45] at (-1.3/1.3,3/1.3) {\small $r=r_{-}$};

\node[rotate=45] at (1.3/1.3,-3/1.3) {\small $r=r_{-}$};	
\node[rotate=-45] at (-1.3/1.3,-3/1.3) {\small $r=r_{-}$};

\node[rotate=45] at (-1.3/1.3,-1/1.3) {\small $r=r_{+}$};	
\node[rotate=-45] at (1.3/1.3,-1/1.3) {\small $r=r_{+}$};

\node[rotate=90] at (-2.3/1.3,3/1.3) {\small $r=0$};
\node[rotate=90] at (2.3/1.3,3/1.3) {\small $r=0$};

\node at (0,-4.6/1.2) {\small $\text{(b)\textbf{ With island}}$};	

\node at (1.5/1.3,1/1.3) {$a_{+}$};	
\draw[blue,fill= blue] (1.5/1.3,0.75/1.3) circle (.035cm);

\node at (-1.5/1.3,1/1.3) {$a_{-}$};	
\draw[blue,fill= blue] (-1.5/1.3,0.75/1.3) circle (.035cm);

\draw[orange] (1.5/1.3,0.75/1.3) .. controls (0, 0.2) .. (-1.5/1.3,0.75/1.3);

\node[rotate=90] at (-2.2/1.3,0) {\small $r=\infty$};	
\node[rotate=90] at (2.2/1.3,0) {\small $r=\infty$};
\node[green] at (3/1.3,-0.1/1.3) {$\mathcal{R}_{+}$};
\node[green] at (-3/1.3,-0.1/1.3) {$\mathcal{R}_{-}$};	
\end{scope}
\end{tikzpicture}
\caption{ The Penrose diagram of the non-extremal charged black planes without island (a) and with island (b) in the presence of a flat thermal bath (red lines) \cite{Ling:2020laa}.\label{fig2} }
\end{figure}

Let us first focus on the entanglement entropy of the Hawking radiation in the absence of the island. Actually, in the early time of the Hawking radiation, there are very few Hawking quanta in the interior of black holes, so the island configuration might not be necessary. The entanglement entropy of Hawking radiation observed from a distant observer can be congruously  described by the s-wave approximation \cite{Hashimoto:2020cas}. Under this approximation, the dynamics of the radiation is effectively a 2-dimensional CFT. Hence, we use the expression of von Neumann entropy for the 2-dimensional CFT to estimate entanglement entropy in our 4-dimensional black hole setup. Therefore, the finite part of the entanglement entropy of the Hawking radiation is given by\footnote{ Note that the UV-divergent part of the entanglement entropy can be absorbed into the renormalized Newton constant \cite{Susskind:1994sm}. The same outcome may be fulfilled by introducing a
generic lower bound cutoff scale in analogy to vacuum energy integrals in order to avoid divergences at infinity. However, we here ignore the cutoff
dependence of the entanglement entropy as we only care about how it scales with the subsystem, see Eq. (\ref{GEformula2}). It is worth mentioning that the UV divergence associated with the endpoints in the bath gets canceled through the area and the matter entropy contributions in the generalized entropy \eqref{GEformula2}.}
 \cite{Hashimoto:2020cas,Calabrese:2009qy,Calabrese:2004eu},
\begin{equation}\label{EEwithoutI}
S( \mathcal{R})=\frac{c}{3} \log \Big[d(b_{+},b_{-})\Big]
\end{equation}
where $d(b_{+},b_{-})$ is the distance between the boundaries $b_{+}(t_{b},b)$ and $b_{-}(-t_{b}+ i \pi \kappa_{+}^{-1})$ of the radiation regions in the right and the left wedge of our Penrose diagrams as illustrated in part (a) of Figs. \ref{fig1}, \ref{fig2}, and \ref{fig3} for the neutral, non-extermal, and extremal black plane, respectively. In all panels, the 4-dimensional flat spacetimes
are used as auxiliary thermal baths to be glued to the both sides of the two-sided planar black hole \cite{Almheiri:2019yqk,Ling:2020laa}.
 Due to thermal equilibrium, the temperature of faraway thermal bath is the Hawking temperature defined in Eq. \eqref{tem}.
 The geodesic distance between two points
 is obtained through
\begin{eqnarray}\label{dfunction}
\nonumber d^2(O_{1},O_{2}) &=&\mathcal{F}(O_{1}) \mathcal{F}(O_{2}) \Big[(U(O_{2})-U(O_{1}))\\
& \times & (V(O_{1})
-V(O_{2}))\Big]
\end{eqnarray}
where $\mathcal{F}$ is the conformal factor and $(U,V)$ are Kruskal coordinates (see Appendix \ref{KScoor} in more detail.). Now by taking together Eqs. \eqref{metricelement} and  \eqref{CFfun}, we obtain that the entropy without island has the following expression,
\begin{eqnarray}\label{EntwithoutI}
\nonumber S(\mathcal R)&=&\frac{c}{6}\log\Big[4 \kappa_{+}^{-2} e^{2 \kappa_{+} r^*(b)}\cosh ^{2}(\kappa_{+}t_{b}) \mathcal{F}(b)^2\Big]\\
 &=&\frac{c}{6}\log\Big[4 b^2 \kappa_{+}^{-2} \cosh ^{2}(\kappa_{+}t_{b}) f(b)\Big]
\end{eqnarray}
At the late-time limit ($t_{b} \to \infty$), writing $\cosh \kappa_{+} t \sim e^{\kappa_{+} t}$, this entropy is linear in time, that is,
\begin{equation}\label{NoIEn}
S(\mathcal R) \simeq \frac{c}{3} \kappa_{+} t \simeq \frac{2 \pi c}{3} T t
\end{equation}
This finding implies that entanglement entropy of the Hawking radiation in the absence of the island surface is increasing linearly with time and becomes infinite at late times which results in the information paradox.

\subsection{Entropy with islands}

Now we compute the entanglement entropy of the Hawking radiation, however upon taking into account the island contribution under the s-wave approximation \cite{Hashimoto:2020cas}. Here, we assume the case when the boundary $r=b$ of the entanglement region $R$ is far away from the horizon, $b \gg r_{+}$, so that the s-wave approximation is valid. Thus, we can use the entanglement entropy of the quantum matter in the overall region of the
union of the radiation and island and the island region \cite{Hashimoto:2020cas}, that is,
\begin{eqnarray}
S_{\text{m}}= \frac{c}{3} \log\Big[\frac{d(a_{+},a_{-}) d(b_{+},b_{-})d(a_{+},b_{+})d(a_{-},b_{-})}{d(a_{+},b_{-})d(a_{-},b_{+})}\Big]\;\;\;\;\;
\end{eqnarray}
Since the distance between the right wedge and the left wedge is very large at late times, one can assume,
\begin{equation}
d(a_{+},a_{-}) \approx d(b_{+},b_{-}) \approx d(a_{\pm},b_{\mp}) \gg d(a_{\pm},b_{\pm})
\end{equation}
The entanglement entropy of the matter, therefore, is approximated as follows,
\begin{equation}
S_{\text{m}}(\mathcal R \cup \mathcal I) \approx \frac{c}{3} \log\Big[d(a_{+},b_{+}) d(a_{-},b_{-})\Big]
\end{equation}
where $a_{+}= (t_{a},a)$ and $a_{-} = (-t_{a} - i \pi /\kappa_{+}, a)$ denote the boundary of the island as shown in part (b) of Figs. \ref{fig1} and \ref{fig2}.
Finally, the generalized entropy is written as the sum of the above entropy of the quantum matter and the area term with respect to boundaries of the island,
\begin{equation}
S_{\text{gen}}=4 \pi \text{Area}(\partial \mathcal{I})+S_{\text{m}}(\mathcal{R}\cup \mathcal{I})
\end{equation}
In Kruskal-Szekeres coordinates as Appendix \ref{KScoor} presents, the generalized entropy can be expressed as,
\begin{eqnarray}\label{Entropy1}
\nonumber S_{\text{gen}}&=&2 \pi a^2+\frac{c}{3} \log\Big[\kappa_{+}^{-2}\Big(e^{2 \kappa_{+} r^{*}(a)}+e^{2 \kappa_{+} r^{*}(b)}\\
\nonumber &-&2e^{\kappa_{+}(r^{*}(a)+r^{*}(b))} \cosh(\kappa_{+}(t_{a}-t_{b}))\Big)\Big]\\
&+& \frac{c}{3} \log\Big[  a b e^{-\kappa(r^{*}(a)+r^{*}(b))} \sqrt{f(a)f(b)}\Big]
\end{eqnarray}
Indeed, to compute the entanglement entropy, one should maximize $S_{\text{gen}}$ in time direction $t=t_{a}$ upon finding the position of an extremizing $S_{gen}$ across all possible Cauchy surfaces.
It is obvious that the generalized entropy is maximazed by $t_{a}=t_{b}$. As a consequence of the above discussion, the time dependent part of the generalized entropy is removed  upon setting $t_{a}=t_{b}$, and the entropy will become a constant at late times. On the other hand,
imposing on Eq. \eqref{Entropy1},  %by plunging $r^{*}(a)$ by
\begin{equation}\label{af2function}
r^{*}(a)=-\frac{1}{2\kappa_{+}}\log\Big[\frac{\mathcal{F}^{2}(a)}{a^2 f(a)}\Big]
\end{equation}
and then extremizing the entropy  with respect to $a$, i.e. $\partial_{a}S_{\text{gen}}=0$ under the assumption $t_{a}=t_{b}$, one can derive the location of the island. Therefore, we have,
\begin{equation}\label{fd}
 2 \pi a+ \frac{c}{3}\frac{\mathcal{F}'(a)}{\mathcal{F}(a)}+\frac{c \mathcal{F}^2(a) \mathcal{H}'(a)}{3a \Big(a f(a)-e^{\kappa_{+} r^{*}(b) \sqrt{f(a) }}\mathcal{F}(a)\Big)}=0
\end{equation}
where $\mathcal{H}(a)=\frac{a^2 f(a)}{\mathcal{F}^2(a)}$.
As we consider the configuration in which the island is located near the event horizon, that is $a =r_{+}+\epsilon$ with $\epsilon \ll 1$ \cite{Lu:2021gmv,Cao:2021ujs}, using the approximation
\begin{eqnarray}
f(a) \approx f'(r_{+})\epsilon+\mathcal{O}(\epsilon^2)
\end{eqnarray}
the solution to Eq.\eqref{fd} is given by,
\begin{equation}\label{asol}
a \approx r_{+}+ \frac{r_{+}^2\kappa_{+}c^2}{36 \pi^2  \mathcal{F}^2(r_{+}) } e^{-2 \kappa_{+} r^{*}(b)}.
\end{equation}
Here $\mathcal{F}^2(r_{+}) = \lim_{ r \to r_{+}} \mathcal{F}^2(r)$ when $c \ll 1$, that is $c\; G_N\ll 1$ upon restoring the Newton constant. This finding is in agreement with the one in \cite{He:2021mst}.  We note that  the conformal factor $\mathcal{F}$ is finite at the horizon.
We now check whether or not the quantum extremal surface exists slightly outside the outer horizon at the late time regime.

\begin{itemize}
\item \textbf{Neutral Cases:}
\end{itemize}
 As one considers $K(X)=X$, in the range $0<\alpha\leq \sqrt{2} r_{+}$, the tortoise coordinate $r^{*}$ is obtained to be,
 \begin{eqnarray}
\nonumber  r^{*}=\int \frac{dr}{r^2 f(r)}&=& \frac{8 r_{+}(3 r_{+}^2-\alpha^2)}{\kappa_{+} \sqrt{\beta}}
\arctan\Big[\frac{2r+r_{+}}{\sqrt{\beta}}\Big]\\
 &-&\frac{4r_{+}^3}{\kappa_{+}} \log\Big(\frac{(r-r_{+})^2}{2(r^2+r r_{+}+r_{+}^2)-\alpha^2}\Big)\nonumber\\
 \end{eqnarray}
 where $\kappa_{+}=\frac{6 r_{+}^2-\alpha^2}{4 r_{+}}$ and $\beta=3 r_{+}^2-2 \alpha^2$. Therefore, the conformal factor \eqref{CFfun} reads,
 \begin{eqnarray}
 \mathcal{F}^2|_{r_{+}} = \frac{(6 r_{+}^2-\alpha^2)^{\frac{3}{2}}}{2 r_{+}} \exp \Big[\frac{\alpha^2-3 r_{+}^2}{r_{+}\sqrt{\beta}} \arctan \Big(\frac{3 r_{+}}{\sqrt{\beta}}\Big)\Big]&&\;\;\;\;\;
\end{eqnarray}
which is finite at the outer horizon and positive. Using Eq. \eqref{asol}, we therefore obtain the location of the island for small value of $\alpha$, as follows,
\begin{eqnarray}
\nonumber a &\approx & r_{+}+\frac{\alpha^2(2 b^2-b r_{+}-r_{+}^2)+6 r_{+}^2(b^2+b r_{+}+r_{+}^2)}{432 \sqrt{3} \pi^2 r_{+} (b-r_{+}) \sqrt{b^2+b r_{+}+r_{+}^2}}\\
&\times & \exp\Big[\frac{\pi- 3 \arctan\Big(\frac{2 b+ r_{+}}{\sqrt{3} r_{+}}\Big)}{\sqrt{3}}\Big]
\end{eqnarray}
Clearly, the result confirms that the quantum extremal surface exists slightly outside the outer horizon.
\begin{itemize}
\item \textbf{Charged Cases:}
\end{itemize}
As mentioned in the previous section, in the range $\sqrt{2} r_{+}<\alpha<\sqrt{6} r_{+}$ for $K(X)=X$, the inner horizon appears before the outer horizon.  Therefore, we obtain the following expression for $r^{*}$,
\begin{eqnarray}
\nonumber r^{*}&=&\frac{1}{2 \kappa_{+}}\log(r-r_{+})+\frac{1}{2 \kappa_{-}}\log(r-r_{-})\\
&+&\frac{(r_{+}-r_{-})^2 (r_{+}+r_{-})}{4 \kappa_{+} \kappa_{-} r_{-}r_{+}}\log(r+r_{+}+r_{-})
\end{eqnarray}
where,
\begin{equation}
\kappa_{\pm}=\frac{(r_{\pm}-r_{\mp})(2 r_{\pm}+r_{\mp})}{2 r_{\pm}}
\end{equation}
 Thus, the conformal factor is determined such,
\begin{equation}
\mathcal{F}^2(r_{+})=\frac{1}{r_{+}} (r_{+}-r_{-})^{1-\frac{\kappa_{+}}{\kappa_{-}}} (2 r_{+}+r_{-})^{1-\frac{(r_{+}-r_{-})^2 (r_{+}-r_{-})}{2 \kappa_{-} r_{-}r_{+}}}
\end{equation}
Finally, the solution to $a$ is obtained as,
\begin{eqnarray}
\nonumber a & \approx & r_{+}+\frac{\kappa_{+} r_{+}^3}{36 \pi^2 (b-r_{+})(r_{+}-r_{-})(2 r_{+}+r_{-})}\\
&\times & \Big(\frac{r_{+}-r_{-}}{b-r_{-}}\Big)^{\frac{\kappa_{+}}{\kappa_{-}}}\Big(\frac{2 r_{+}+r_{-}}{b+r_{+}+r_{-}}\Big)^{\frac{(r_{+}-r_{-})^2(r_{+}+r_{-})}{2 \kappa_{-} r_{+} r_{-}}}
\end{eqnarray}
It is obvious that the island position is outside the outer event horizon.
In addition, for the model $K(X)=X^2$, the location of the island is obtained to be,

\begin{widetext}
\begin{eqnarray}\label{ax2}
&& a \simeq r_{+}+\frac{2 r_{+}^2 \kappa_{+}^2}{36 \pi^2(b-r_{+})}  \Big((b-r_{-})(r_{+}-r_{-})\Big)^{-\frac{\kappa_{+}}{\kappa_{-}}} \Big(\frac{2\kappa_{+} r_{+}^2 (b^2+r_{+}^2+r_{+}r_{-}+r_{-}^2+b(r_{+}+r_{-}))}{r_{+}-r_{-}}\Big)^{-\frac{(r_{+}-r_{-})^2(r_{+}+r_{-})^3}{4 \kappa_{-} r_{+}^2 r_{-}^2}} \\
\nonumber && \exp \Big[\frac{(r_{+}-r_{-})(\kappa_{+} r_{+}^4+r_{-}\Big(2 r_{+}^4+2 r_{+}^3 r_{-}-2 r_{+}^2 r_{-}^2-(\kappa_{-}+2 r_{+}) r_{-}^3\Big))}{ \kappa_{-} r_{-}^2 r_{+}^2 \sqrt{\gamma}} \Big(\arctan \Big(\frac{2 b+ r_{+}+r_{-}}{\sqrt{\gamma}}\Big)-\arctan \Big(\frac{ 3 r_{+}+r_{-}}{\sqrt{\gamma}}\Big)\Big) \Big]
\end{eqnarray}
\end{widetext}
where $\gamma= \frac{2 \kappa_{+} r_{+}^2}{r_{+}-r_{-}}+2 r_{-}^2$ and,
\begin{equation}
\kappa_{\pm}=\frac{(r_{\pm}-r_{\mp})}{4 \pi r_{\pm}^2}(3 r_{\pm}^2+2 r_{+}r_{-}+r_{\mp}^2)
\end{equation}
 Finally, the $K(X)=X^{3}$ model results in,

\begin{eqnarray}\label{ax3}
\nonumber a &\simeq & r_{+}+\frac{r_{+} \sqrt{b^2+b r_{+}+r_{+}^2}}{36 \sqrt{3} \pi^2 (b-r_{+})}\Big(\frac{(r_{+}-r_{-})^3 (b^2+ b r_{-}+r_{-}^2)}{3 \kappa_{+}(b-r_{-})^2 r_{+}^2 }\Big)\\
\nonumber &\times & \exp \Big[\frac{1}{\sqrt{3} r_{+}^2}\Big(\pi r_{+}^2-3 r_{+}^2 \arctan \Big[\frac{2 b+ r_{+}}{\sqrt{3} r_{+}}\Big]\\
&+& 3 r_{-}^2 \Big(\arctan \Big[\frac{2 b+ r_{-}}{\sqrt{3} r_{-}}\Big]-\arctan \Big[\frac{2 r_{+}+ r_{-}}{\sqrt{3} r_{-}}\Big]\Big)\Big)\Big]
\end{eqnarray}
where,
\begin{equation}
\kappa_{\pm}=\frac{3}{2 r_{\pm}^2} \Big(r_{\pm}^3-r_{\mp}^3\Big)
\end{equation}
It is straightforward, using Eqs. \eqref{ax2} and \eqref{ax3}, to verify that the island for both $X^2$ and $X^3$ is placed out of the outer even horizon.

Now, substituting the approximate solution \eqref{asol} into \eqref{Entropy1}, the late-time radiation entropy with island is expressed by,
\begin{eqnarray}\label{SIE}
 &S(\mathcal R)&=2 \pi r_{+}^2+\frac{c}{3} \log\Big[\frac{e^{2 \kappa r^{*}(b)}}{\kappa_{+}^2} \mathcal{F}(b) \mathcal{F}(r_{+}) \Big]\\
\nonumber &+& \frac{c^2 r_{+}^3 \kappa_{+}}{18 \pi \mathcal{F}(r_{+})^2} e^{-2 \kappa_{+} r^{*}(b)}+ \mathcal{O}(c^3)\approx 2 S_{BH}+ \mathcal{O}(c)
\end{eqnarray}
As we observe, the main contribution to entanglement entropy at late times is from the area
term, that being twice the Bekenstein–Hawking entropy of black planes. The present results substantiates the previous findings in the literature for other black holes \textit{e.g.} \cite{ Hashimoto:2020cas,Lu:2021gmv,Yu:2021cgi,Ahn:2021chg,He:2021mst}.

\subsection{Page Times and Scrambling Times}
In early times and in the absence of the island, as we have observed, the entanglement entropy grows linearly in time. Nevertheless, at late times, the island appears near the outer horizon and the resultant entanglement entropy becomes constant, namely $2 S_{BH}$. Equating the asymptotic constant value of the entropy \eqref{SIE} with the entropy without the island \eqref{NoIEn}, the Page time for our eternal black planes is obtained to be,
\begin{equation}\label{pageex}
c\, t_{\text{Page}} = \frac{3 S_{\text{BH}}}{ \pi T_{H}}
\end{equation}
 The above finding implies that the island arises around the Page time which itself is proportional to twice the Bekenstein-Hawking entropy of the black hole. If the massive gravity parameter $\alpha$ is set to be close enough to the extremal limit \eqref{alphiex}, so that the inner and outer horizons are sufficiently close each other, the Page time is estimated as,
 \begin{equation}\label{pageapp}
c\, t_{\rm Page} \approx \frac{4 \pi r_{+}^2}{n (r_{+}-r_{-})} + \frac{4 \pi r_{+}}{3}
\end{equation}

Table \ref{tab1} presents great agreements between the exact and approximation results for the Page time. Notice that $r_{-}$ is a function of $\alpha$ and $r_{+}$ according to the equation $Q(r_{-})=0$. Clearly, the Page time does decrease upon increasing the power of $K(X)=X^{n}$ models. It means that when the higher orders of massive gravity are added to the theory we are getting the Page curves at earlier time. Furthermore, for neutral black plane cases, in the small massive gravity parameter limit, \textit{i.e.} $\alpha \ll 1$, the Page time is obtained in terms of $\alpha$ as,
\begin{equation}
c\, t_{\rm Page}= 4 \pi r_{+}+ \frac{2 \pi r_{+}}{3} \alpha^{2n}+ \mathcal{O}(\alpha^{4 n})
\end{equation}
whose leading term is the Page time for the neutral black plane in the context of Einstein gravity, corresponding to the massless graviton with unbroken diffeomorphism, and the subleading terms capture the contributions of the massive gravity. Since $n>0$, this implies that the mass deformation makes the evaporation of the neutral black plane reaching more slowly the Page time compared to the original theory with massless gravitons.

 As a consequence, the radiation entropy of the radiation increases linearly with time at early time, during which no island is formed. Around the Page time, the island can be formed near the horizon of the black hole and the entropy takes a nearly constant value which is twice the thermal entropy of the black plane.\\\\

 \begin{table}
 \begin{tabular}{l*{6}{c}r}
Model              & $X$ & $X^2$ & $X^3$ & $X^4$ & $X^5$  & $X^{6}$ \\
\hline
$c \, t_{\rm Page}^{\rm Exact} $ & 128.886 & 67.745 & 47.43 & 37.335 & 31.317 & 27.339  \\
$c \, t_{\rm Page}^{\rm Approx} $ & 128.740 & 67.567 & 47.150 & 36.923 & 30.772 & 26.660 \\
%$t_{\rm scr}$            & 25.237 & 13.943 & 10.233 & 8.419 &  7.362 & 6.682  \\
%Benfica           & 6 & 2 & 1 & 3 &  7 & 8 &  7  \\
%FC Copenhagen     & 6 & 2 & 1 & 3 &  5 & 8 &  7  \\
\end{tabular}
\caption{A comparison between the exact \eqref{pageex} and the approximated \eqref{pageapp} values of the Page time  for various holographic massive gravity models in charged black plane cases as one takes $\alpha=0.95 \alpha_{\rm ext}$ to be near the extremal limit and $r_{+}=1$. \label{tab1}}
 \end{table}

 We now discuss the scrambling times of the planar black holes in the massive gravity models of the present study. As mentioned before, in our asymptotically AdS black hole cases, we couple a flat bath at the boundary of the spacetime. Therefore,
  the scrambling time, $t_{\rm scr}$, is defined to be the minimum time we can recover the information bits, after sending them from the boundary (or the flat bath) to inside the black hole, in the form of the Hawking radiation. If the observer sends the information from the boundary at $r=b$ to the black hole, then the time of arrival at the island surface $r=a$ is $t_{\rm scr}=r^{*}(b)-r^{*}(a)$.
To find this time, we first need to obtain the tortoise coordinate $r^{*}(b)$ from Eq. \eqref{asol} by redefining $\epsilon=a-r_{+}$. Thus,
\begin{equation}
r^{*}(b) \sim \frac{1}{2 \kappa_{+}} \log \Big[\frac{r_{+}^2 c^2 \kappa_{+}}{36 \pi^2 \mathcal{F}^2(r_{+}) \epsilon}\Big]
\end{equation}
 The above expression together with Eq. \eqref{af2function}, one finds the scrambling time to be,
\begin{eqnarray}
t_{\rm scr}=r^{*}(b)-r^{*}(a) \simeq \frac{1}{2 \kappa_{+}} \log \Big[\frac{c^2 r_{+}^2 \kappa_{+}}{a^2 f(a)\epsilon} \Big(\frac{\mathcal{F}(a)}{\mathcal{F}(r_{+})}\Big)^2\Big]&&\;\;\;\;\;
\end{eqnarray}
Since the  island is located near and outside the event horizon, i.e. $a =r_{+}+\epsilon$ with $\epsilon \ll 1$, we can expand the above relation as follows,
\begin{eqnarray}
\nonumber t_{scr} & \simeq & \frac{1}{2 \kappa_{+}} \log \Big[\frac{ c^2 \kappa_{+}}{f'(r_{+}) \epsilon^2}\Big] = \frac{1}{2 \kappa_{+}} \log\Big(r_{+}^2\Big)\\
&+& \text{subleading trems}
\simeq  \frac{1}{T_{H}} \log \Big(S_{BH}\Big)
\end{eqnarray}
In the above expression, we assumed that $c \sim \epsilon \ll S_{BH}$, that is, the central charge is much smaller than the Bekenstein-Hawking
entropy. Clearly, the scrambling time  logarithmically smaller than the life time of the black hole.  This result is in agreement with that one derived in Refs. \cite{Sekino:2008he,Hayden:2007cs}.
Similar to the Page time, for the charged cases with $\alpha \sim \alpha_{ext}$, the scrambling  time will be small at large $n$ as well.
%%%%%%%%%%%%%%%%%%%%%%%%%%%%%%%%%

\section{The Extremal Cases}\label{sec4}
\begin{figure}
\begin{center}
\begin{tikzpicture}[line width=1.5 pt, scale=0.9]
	\draw[vector] (0,4/1.3)--(0,-4/1.3);
	\draw[dashed] (2/1.3,-2/1.3)--(0,0);
	\draw[dashed] (2/1.3,2/1.3)--(0,0);
	\draw[red] (4/1.3,0) -- (2/1.3,-2/1.3)--(0,-4/1.3);
	\draw[red] (4/1.3,0) -- (2/1.3,2/1.3)--(0,4/1.3);
	\draw[black] (2/1.3,2/1.3) -- (2/1.3,-2/1.3);
\node at (2.5/1.3,1/1.3) {$b_{+}$};	
\draw[blue,fill= blue] (2.5/1.3,0.75/1.3) circle (.035cm);
\draw[green,dotted] (4/1.3, 0) .. controls (3/1.3, 0.2/1.3) .. (2.5/1.3,0.75/1.3 );

\node[green] at (3/1.3,-0.1/1.3) {$\mathcal{R}_{+}$};

\node[rotate=90] at (-0.3,0) {\small $r=0$};

\node[rotate=45] at (1.3/1.3,1/1.3) {\small $r=r_{e}$};	
\node[rotate=-45] at (1.3/1.3,-1/1.3) {\small $r=r_{e}$};

\node[rotate=90] at (2.2/1.3,0) {\small $r=\infty$};

\node at (1,-4.6/1.2) {\small $\text{(a) \textbf{Without island}}$};	
\begin{scope}[shift={(3.6,0)}]
\draw[vector] (0,4/1.3)--(0,-4/1.3);
	\draw[dashed] (2/1.3,-2/1.3)--(0,0);
	\draw[dashed] (2/1.3,2/1.3)--(0,0);
	\draw[red] (4/1.3,0) -- (2/1.3,-2/1.3)--(0,-4/1.3);
	\draw[red] (4/1.3,0) -- (2/1.3,2/1.3)--(0,4/1.3);
	\draw[black] (2/1.3,2/1.3) -- (2/1.3,-2/1.3);
\node at (2.5/1.3,1/1.3) {$b_{+}$};	
\draw[blue,fill= blue] (2.5/1.3,0.75/1.3) circle (.035cm);
\draw[green,dotted] (4/1.3, 0) .. controls (3/1.3, 0.2/1.3) .. (2.5/1.3,0.75/1.3 );

\node[green] at (3/1.3,-0.1/1.3) {$\mathcal{R}_{+}$};

\node[rotate=90] at (-0.3,0) {\small $r=0$};

\node[rotate=-45] at (1.3/1.3,-1/1.3) {\small $r=r_{e}$};

\node at (1,-4.6/1.2) {\small $\text{(b)\textbf{ With island}}$};	

\node at (1.5/1.3,1/1.3) {$a_{+}$};	
\draw[blue,fill= blue] (1.5/1.3,0.75/1.3) circle (.035cm);

\draw[orange] (1.5/1.3,0.75/1.3) .. controls (0.5, 0.3) .. (0,0.75/1.3);

\node[rotate=90] at (2.2/1.3,0) {\small $r=\infty$};

\end{scope}
\end{tikzpicture}
\caption{The Penrose diagram of the extremal black planes without island (a) and with island (b) in the presence of a flat thermal bath (red lines).\label{fig3} }
\end{center}
\end{figure}
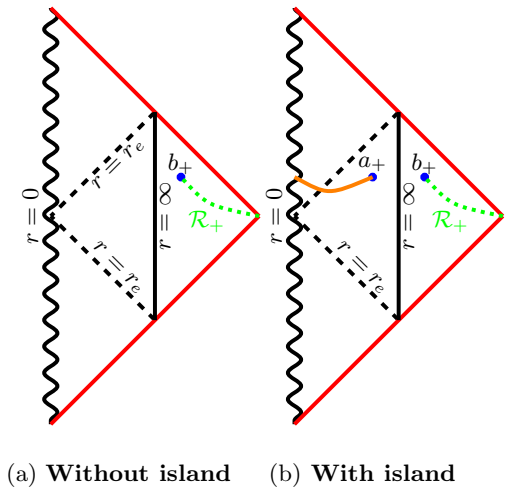

In comparison to the non-extemal black plane, within the island framework, we must perform an independent analysis for the entanglement entropy in the extremal black plane. For instance, the radiation region for the extremal charged black plane is only given by one region $R_{+}$ as the Penrose diagram illustrated in Fig. \ref{fig3}. As shown in this Penrose diagram, the Cauchy surface including $b_{+}=(t_{b},b)$ touches the singularity at $b_{0}=(t_{b},0)$. In the absence of the island, from Eqs. \eqref{EEwithoutI} and \eqref{dfunction},  the finite part of the entanglement entropy reads
\begin{eqnarray}\label{EEEter}
&&   S(\mathcal R) \sim  \log \Big[\mathcal{F}(b)\mathcal{F}(0)(U(b)-U(0))(V(0)-V(b)) \Big]\hspace{0.5cm}
\end{eqnarray}
One can easily check that the conformal factor for the extremal metric function \eqref{frmre} is divergent at the singularity ($r=0$) as
\begin{equation}
\mathcal{F}|_{r \to 0} \sim \Bigg\{\begin{matrix}
 r^{-1}& \text{for} & n<\frac{3}{2} \\
 r^{2-2n}& \text{for} & n>\frac{3}{2}
\end{matrix}
\end{equation}
As a consequence of this fact, the entanglement entropy in the absence of the island \eqref{EEEter} is ill-defined for such a black planar.
The same result has been reported for other extremal charged black hole geometries, see \textit{e.g.} \cite{Ahn:2021chg,Kim:2021gzd,Cao:2021ujs}.
On the other hand, to compute the entanglement entropy in the presence of the island configuration, we first need to rewrite the matter part of Eq. \eqref{Entropy1} as follows,
 \begin{eqnarray}\label{EXEntropy1}
 S_{\text{m}}&=&\frac{c}{3} \log\Big[\kappa_{+}^{-2} \Big(\cosh(\kappa_{+}(r^{*}(a)-r^{*}(b)))\\
\nonumber &-& 2 \cosh(\kappa_{+} (t_{a}-t_{b}))\Big)\Big]+\frac{c}{3} \log\Big[  a b  \sqrt{f(a)f(b)}\Big]
\end{eqnarray}
In addition, equations \ref{frmrp} and \ref{tem} imply that the surface gravity $\kappa_{+} \to 0$ in extremal cases. Therefore, by taking the limit $\kappa_{+} \to 0$ from Eq. \eqref{EXEntropy1}, the generalized entropy for the extremal case \cite{Kim:2021gzd} is written as,
\begin{eqnarray}
&& S_{\text{gen}}=  \pi a^2+ \lim_{\kappa_{+} \to 0} S_{\text{m}}
\approx  \pi a^2 \\
\nonumber &&+ \frac{c}{3}\log\Big[ a b  \sqrt{f(a)f(b)} \Big((r^{*}(a)-r^{*}(b))^2-(t_{a}-t_{b})^2\Big)\Big]
\end{eqnarray}
in which the first contribution comes form area term. In the above relation, we have used the approximation $\cosh(x) \approx 1+x^2/2$. It is trivial to verify that $t_{a}$ equals to $t_{b}$, when we extremise $S_{\text{gen}}$ with respect to $t_{a}$. Hence, we have,
\begin{eqnarray}\label{EEntropy}
S_{\text{gen}}= \pi a^2+ \frac{c}{3}\log\Big[ a b  \sqrt{f(a)f(b)} \Big((r^{*}(a)-r^{*}(b))^2\Big)\Big]&& \;\;\;\;\;\;\;\;
\end{eqnarray}
Now, by varying $S_{\text{gen}}$ with respect to $a$, one obtains the following algebraic equation determining
the location of the island,
\begin{equation}\label{aloc2}
2 \pi a+ \frac{c}{3 a}+ \frac{c}{6} \frac{f'(a)}{f(a)}+\frac{2 c}{3 a^2 f(a)} \Big(r^{*}(a)-r^{*}(b)\Big)^{-1}=0
\end{equation}
Similar to the non-extremal case, we use the ansatz that the island is located near the event horizon, that is $a =r_{e}+\epsilon$ with $\epsilon \ll 1$, together with the following approximations,
\begin{eqnarray}
\nonumber f(a) &\approx & \frac{1}{2}f''(r_{e})\epsilon^2+\mathcal{O}(\epsilon^3)\\
r^{*}(a) &\approx & -\frac{2}{r_{e}^2 f''(r_{e})\epsilon}
\end{eqnarray}
Hence, the solution to Eq.\eqref{aloc2} reads,
\begin{equation}\label{asol1}
a \approx r_{e}+ \frac{c}{6 \pi r_{e}}-\frac{c^2 \Big(5 +r_{e}^3 f''(r_{e})r^{*}(b)\Big)}{36 \pi^2 r_{e}^3}+ \mathcal{O}(c^3)
\end{equation}
For instance, we take the metric form the \eqref{emetric} for the $K(X) \in \{X,X^2,X^3\}$ model in the tortoise coordinate,
\begin{eqnarray}
r_{X}^{*}(b)&=& -\frac{1}{3 (b-r_{e})}+\frac{2}{9 r_{e}} \log\Big[\frac{1- \frac{r_{e}}{b}}{1+2 \frac{r_{e}}{b}}\Big]\\
\nonumber r_{X^{2}}^{*}(b)&=& -\frac{1}{6 (b-r_{e})}+\frac{1}{9 r_{e}} \log\Big[\frac{(1- \frac{r_{e}}{b})^2}{1+2 \frac{r_{e}}{b}+3 \Big(\frac{r_{e}}{b}\Big)^2}\Big]\\
 &+& \frac{7}{18 \sqrt{2} r_{e}}\arctan\Big[\frac{1+\frac{r_{e}}{b}}{\sqrt{2} \frac{r_{e}}{b}}\Big]\\
\nonumber r_{X^{3}}^{*}(b)&=& -\frac{1}{b (1-\Big(\frac{r_{e}}{b}\Big)^3)}+\frac{1}{9 r_{e}} \log\Big[\frac{(1- \frac{r_{e}}{b})^2}{1+ \frac{r_{e}}{b}+ \Big(\frac{r_{e}}{b}\Big)^2}\Big]\\
 &+& \frac{2}{3 \sqrt{3} r_{e}}\arctan\Big[\frac{2+\frac{r_{e}}{b}}{\sqrt{3} \frac{r_{e}}{b}}\Big]
\end{eqnarray}

Since we assume that the cutoff surface locates far from the horizon, $b \gg r_{e}$, one obtains,
\begin{eqnarray}
r_{X}^{*}(b) \approx
r_{X^2}^{*}(b) \approx
r_{X^3}^{*}(b) \approx -\frac{1}{b}
\end{eqnarray}
Therefore, the island location for the above examples is determined to be,
\begin{equation}
a \approx r_{e}+ \frac{c}{6 \pi r_{e}}+ \mathcal{O}(c^2)
\end{equation}
Interestingly, Eq. \eqref{asol1} can also reproduce the results for both the extremal Reissner–Nordstr$\ddot{\text{o}}$m and the extremal Hayward black holes in \cite{Kim:2021gzd}. Indeed, our findings reveal the universality of island position which is outside the horizon $r_{e}$ for all extremal configurations.
Generally, by substituting the approximate solution \eqref{asol1} into \eqref{EEntropy}, the late-time radiation entropy with island reads,
\begin{eqnarray}
\nonumber S(\mathcal R)&\approx & \pi r_{e}^2+ \frac{c}{3} \log\Big[\frac{12 \pi \sqrt{2} b}{ c  r_{e}^2 f''(r_{e})^2} \sqrt{f(b) f''(r_{e})}\Big]\\
\nonumber &+& \frac{c^2\Big(5 +r_{e}^3 r^{*}(b) f''(r_{e})\Big)}{18 \pi r_{e}^2} +\mathcal{O}(c^3)\\
 &\approx & S_{BH}+ \mathcal{O}(c \log c)
\end{eqnarray}
The above finding for the late-time asymptotics of the entanglement entropy of the Hawking radiation is in line with the previous results \cite{Ahn:2021chg,Kim:2021gzd,Cao:2021ujs}. That is, the radiation entropy of the extremal charged planar black hole does approximate a finite constant at the late times, like the associate neutral and non-extremal charged black hole solutions. However, we find that the asymptotic constant value for the entanglement entropy for the extremal charged black plane is approximately the Bekenstein-Hawking entropy, rather than the double of the Bekenstein-Hawking entropy for the non-extremal and neutral cases. The reason behind this refers to their causal structure. More precisely, the Penrose diagram of the neutral and non-extremal charged black plane (see Figs. \ref{fig1} and \ref{fig2}) is the two-sided geometry, whereas the Penrose diagram of the extremal charged black plane is the one-sided geometry as shown in Fig. \ref{fig3}. Moreover, the Page time was not found in the extremal cases, given the absence of the linear growth of the entanglement entropy in the early times.

\section{Conclusion and Discussion}\label{sec5}
The present study has investigated the entanglement entropy of the Hawking radiation from the black planes in holographic massive gravity. One main distinction of the black planes we explored here from common neutral black planes is the appearance of the inner horizon due to the induced mass of the graviton. To calculate the entanglement entropy, we assumed that the eternal black planes with/without an island are in thermal equilibrium with a heat bath.

In the absence of the island, the entanglement entropy of the Hawking radiation grows linearly with time forever in neutral and non-extermal cases. This however implies that the black plane spacetime evolves from a pure state to a mixed state. Nevertheless, due to the different causal structure for extremal black planes, one could not see a linear growth of the entanglement entropy as such in the early times.

We have observed, on the other hand, that for neutral and non-extremal black planes, the entanglement entropy of the Hawking radiation is bounded by a constant value
that is twice the Bekenstein–Hawking entropy ($S(\mathcal R) \sim 2 S_{BH}$) in the presence of the island. However, in the extremal case, this constant equals Bekenstein-Hawking entropy ($S(\mathcal R) \sim S_{BH}$). This difference seems natural becauase the causal structure of the extremal case forms one-side black plane.

Moreover, a universal feature for the existence of the quantum extremal surface in the extremal black planes has been found in our study. More precisely, for $c \ll 1$, there must be a quantum extremal surface situated in the near-horizon region outside the extremal black plane, see Eq. \eqref{asol1}. This result holds independently of the model specifying the chosen parameters. Moreover, we have found for the neutral and non-extremal black planes that the boundary of the island is located slightly outside the outer horizon.

Finally, we have demonstrated how the Page time is affected by the holographic massive gravity deformations. To
be more explicit, for either neutral solutions at the small mass-generating parameter $\alpha \to 0$, or for charged black planes with $\alpha \sim \alpha_{\rm ext}$, the Page time can happen at earlier times.

Future studies can be carried out in order to investigate the entanglement entropy of the Hawking radiation form hairy black hole solutions in massive gravity, \textit{e.g.}  \cite{Mansoori:2021wxf,Mirjalali:2022wrg}.

\section*{Acknowledgments}
We are grateful to Hyun-Sik Jeong and Matteo Baggioli for their valuable suggestions. Further, we acknowledge the anonymous referees for their valuable comments.

\appendix
\section{Thermodynamic quantities }\label{sec: sec6}
 In this appendix, we analyze the thermodynamics of the black plane solution \eqref{metricelement} by applying the solution phase space method proposed in \cite{Hajian:2014twa,Hajian:2015xlp,Ghodrati:2016vvf,Chernyavsky:2017xwm,Hajian:2021hje}. To do this, one first needs to find the surface term $\Theta$ which is obtained through the variation of the Lagrangian \eqref{action1} with respect to the dynamical field, namely,
 \begin{eqnarray}
\nonumber  \delta \Big(\sqrt{-g} \mathcal{L}\Big)&=& \sqrt{-g} \Big[E_{g}^{\alpha \beta} \delta g_{\alpha \beta}+\sum_{I=1}^{2}E_{\Phi^{I}} \delta \Phi^{I} \Big]\\
 &+& \sqrt{-g} \nabla_{\mu} \Theta^{\mu}
 \end{eqnarray}
 where $E_{i}s$ are the equations of motion (EOM) which are given by,
\begin{eqnarray}
\nonumber E_{g}^{\alpha \beta}&=&-\frac{1}{16 \pi G_{N}} \Big[R^{\alpha \beta}-\frac{1}{2} R g^{\alpha \beta}-\frac{3}{L^2} g^{\alpha \beta}+K(X)g^{\alpha \beta}\\
\nonumber &-& K'(X) \sum_{I=1}^2 \nabla^{\alpha} \Phi^{I} \nabla^{\beta} \Phi^{I}\Big]\\
\nonumber E_{\Phi^{I}}&=&\frac{1}{16 \pi G_{N}} \nabla^{\alpha}\Big(K'(X) \nabla_{\alpha} \Phi^{I}\Big)\\
\nonumber \Theta^{\beta}&=&\frac{1}{16 \pi G_{N}} \Big[\nabla_{\alpha} \delta g^{\alpha \beta}-\nabla^{\beta} \delta g_{\alpha}^{\alpha}-K'(X)\sum_{I=1}^{2}\nabla^{\beta} \Phi^{I} \delta \Phi^{I}  \Big]
\end{eqnarray}
 Then, we need to find the Noether charge $Q^{\mu \nu}$ associated with the diffeomorphism generator $\xi^{\mu}$ from the Noether current $\mathcal{J}^{\mu}=\Theta_{\mu} \Big(\delta_{\xi} g,\delta_{\xi} \Phi^{I}\Big)-\xi^{\mu} \mathcal{L}$. It means that $Q^{\mu \nu}$ finally reads from the on-shell relation $\mathcal{J}=\nabla_{\nu} Q^{\mu \nu}$. Note that the vector $\xi^{\mu}$ acts on the variation fields by Lie derivative as follows,
 \begin{eqnarray}\label{xirela}
\nonumber \delta_{\xi} g_{\alpha \beta}&=&L_{\xi} g_{\alpha \beta}=\bigtriangledown_{\alpha} \xi_{\beta}+\nabla_{\beta} \xi_{\alpha}\\
  \delta_{\xi} \Phi^{I} &=& L_{\xi} \Phi^{I}=\xi^{\alpha} \nabla_{\alpha}\Phi^{I}
 \end{eqnarray}
 Using the above relations and imposing EOMs, the Noether charge $Q^{\mu \nu}$ is given by,
 \begin{equation}
 Q^{\mu \nu}=\frac{1}{16 \pi G_{N}}\Big[\nabla^{\nu} \xi^{\mu}-\nabla^{\mu} \xi^{\nu}\Big]
 \end{equation}
Having this quantity together with surface term enables us to find the other main conserved charge tensor $K^{\mu \nu}$ which can be defined as follows,
\begin{equation}
\sqrt{-g} \mathcal{K}_{\xi}^{\mu \nu}=\delta\Big(\sqrt{-g} Q_{\xi}^{\mu \nu}\Big)-\sqrt{-g} \Big[\xi^{\nu} \Theta_{\xi}^{\mu}-\xi^{\mu} \Theta_{\xi}^{\nu}\Big]
\end{equation}
Notice that to get $\delta\Big(\sqrt{-g} Q_{\xi}^{\mu \nu}\Big)$, the variation of $\xi^{\mu}$ must be vanishing, \textit{i.e.} $\delta \xi^{\mu}=0$. Therefore, the conserved charge tensor is obtained as,
\begin{eqnarray}\label{qq2}
\nonumber \mathcal{K}^{\alpha \beta}_{\xi}&=&-\frac{1}{32 \pi G_{N}}\Big[2 \xi^{\gamma} \nabla^{\alpha} \delta g^{\beta}_{\gamma}-2 \xi^{\gamma} \nabla^{\beta} \delta g^{\alpha}_{\gamma}-2 \xi^{\beta} \nabla^{\alpha} \delta g^{\gamma}_{\gamma}\\
\nonumber &+& 2 \xi^{\alpha} \nabla^{\beta} \delta g^{\gamma}_{\gamma}+\delta g^{\gamma}_{\gamma} \nabla^{\alpha} \xi^{\beta}
-\delta g^{\gamma}_{\gamma} \nabla^{\beta} \xi^{\alpha}+2 \xi^{\beta} \nabla_{\gamma} \delta g^{\alpha \gamma}\\
\nonumber &-& 2 \xi^{\alpha} \nabla_{\gamma} \delta g^{\beta \gamma}+2\delta g^{\beta}_{\gamma} \nabla^{\gamma} \xi^{\alpha}
-2 \delta g^{\alpha}_{\gamma} \nabla^{\gamma} \xi^{\beta}\\
&+&2 K'(X) \sum_{I=1}^{2} \Big( \xi^{\alpha}\delta \Phi^{I} \nabla^{\beta} \Phi^{I}-\xi^{\beta}\delta \Phi^{I} \nabla^{\alpha} \Phi^{I}\Big)\Big]
\end{eqnarray}
On the other hand, based on the phase space method \cite{Hajian:2014twa,Hajian:2015xlp,Ghodrati:2016vvf,Chernyavsky:2017xwm,Hajian:2021hje}, we can always build a manifold $\mathcal{M}$ in the phase space whose points are identified by $\Xi (x^{\mu},p_{j})$ where $p_{j}$ are the independent parameters\footnote{ Note that $\Xi$s are solutions to the EOM in the phase space.}.

Moreover, the tangent space of such a manifold is built upon a subset of perturbations, \cite{Hajian:2014twa}
\begin{equation}
\delta \Xi=\frac{\partial \Xi}{\partial p_{j}} \delta p_{j}.
\end{equation}
Now, combining $\mathcal{K}^{\mu \nu}$  with the parametric variations $\delta \Xi$, we can obtain the conserved charges associated with the exact symmetries $\xi^{\mu}$ of the black plane solutions.\\\\
According to the metric coordinates \eqref{metric12} chosen for a black plane, if one assumes $\delta \Sigma$ to be a surface with constant $(t,r)$, the conserved charge variations for an exact symmetry $\xi^{\mu}$ is defined as \cite{Hajian:2014twa,Hajian:2015xlp,Ghodrati:2016vvf},
\begin{equation}\label{qq3}
\delta H_{\xi}=\int_{\delta \Sigma} \sqrt{- g} \mathcal{K}_{\xi}^{tr} dx dy
\end{equation}
For our case of study, since the dynamical fields $\Xi =\{g^{\mu \nu},\Phi^{I}\}$ depend on the parameters $p_{i} =(r_{+},\alpha)$, the parametric variations appeared in \eqref{qq2} are also given by,
\begin{eqnarray}\label{variation}
\nonumber \delta g_{\xi}^{\alpha \beta}&=&\frac{\partial g_{\xi}^{\alpha \beta}}{\partial \alpha} \delta \alpha+\frac{\partial g_{\xi}^{\alpha \beta}}{\partial r_{+}} \delta r_{+}\\
\delta \Phi^{I}_{\xi}&=& \frac{\partial \Phi^{I}_{\xi}}{\partial \alpha} \delta \alpha+\frac{\partial \Phi^{I}_{\xi}}{\partial r_{+}} \delta r_{+}
\end{eqnarray}
To read out the mass of the black plane solution \eqref{metricelement}, one must choose the Killing vector $\xi^{\mu}=-\partial_{t}$ as the generator to which the mass is
associated. Note that the minus sign has been adopted to make the mass to be positive. By using Eqs. \eqref{qq2}, \eqref{variation}, \eqref{qq3}, and the metric element \eqref{metricelement}, we therefore arrive at,
\begin{equation}
\delta M=\delta H_{-\partial t}=\frac{r_{+}^2}{32 \pi G_{N}} \Big[6-K\Big(\frac{\alpha^2}{r_{+}^2}\Big)\Big]\delta r_{+}
\end{equation}
Hence,
\begin{equation}
M= \frac{1}{32 \pi G_{N}}\int^{r_+} s^2 \Big[6-K\Big(\frac{\alpha^2}{s^2}\Big)\Big] ds
\end{equation}
Note that here we have considered the volume of the surface $\delta \Sigma$ set to be one.
In addition, according to the definition of the surface gravity,
\begin{equation}
\kappa_{+}^2=-\frac{1}{2} \nabla_{\mu} \xi_{\nu}\nabla^{\mu} \xi^{\nu}
\end{equation}
the corresponding temperature is given by,
\begin{equation}
T=\frac{\kappa_{+}}{2 \pi}= \frac{r_{+}^2 f'(r_{+})}{4 \pi}=\frac{r_{+}}{8 \pi } \Big[6-K\Big(\frac{\alpha^2}{r_{+}^2}\Big)\Big]
\end{equation}
The horizon entropy is defined to be the conserved charge associated with,
the horizon killing vector $\xi_{h}$ normalized by the Hawking
temperature $T$. Therefore, by choosing $\xi_{s}=\xi_{h}/T$ with $\xi_{h}=-\partial t$, the entropy at the horizon is obtained to be,
\begin{equation}
\delta S=\delta H_{\xi_{s}}=\frac{4 \pi r_{+}}{16 \pi G_{N}} \delta r_{+}= \delta \Big(\frac{ \pi r_{+}^2}{8 \pi G_{N}}\Big)
\end{equation}
And then,
\begin{equation}
 S= \frac{ \pi r_{+}^2}{8 \pi G_{N}}
\end{equation}
 These quantities satisfy the first law of thermodynamics, that is, $\delta M = T \delta S$.
 %%%
\section{Kruskal-Szekeres Coordinates}\label{KScoor}
A general form of the conformal factor function in the Kruskal-Szekeres coordinate is presented. First of all, we take the following general ansatz for the metric of spacetime,
\begin{equation}\label{metric1}
	ds^2=-r^2 h(r)dt^2+\frac{dr^2}{r^2 f(r)}+r^2(dx^2+dy^2)
\end{equation}
Then, by considering the null geodesic, we can introduce the tortoise coordinate as,
\begin{equation}
	dr_*=\frac{1}{r^2 \sqrt{h(r) f(r)}}
\end{equation}
Now, the line element \eqref{metric1} converts to,
\begin{equation}\label{}
	ds^2=r^2 h(r)(-dt^2+dr_* ^2)+r^2(dx^2+dy^2)
\end{equation}
On the other hand, by defining both ingoing and outgoing radial null coordinates, \textit{i.e.}
 $v=t+r_{*}$ and $u=t-r_{*}$ respectively, the above metric becomes,
\begin{equation}\label{}
	ds^2=-r^2 h(r)du dv+r^2(dx^2+dy^2).
\end{equation}
Now, switching to the new coordinates $(U,V)$ which are defined by,
\begin{eqnarray}
	U=-\kappa_{+}^{-1} e^{\kappa_{+} u}, \hspace{0.5cm} V= \kappa_{+}^{-1} e^{ \kappa_{+} v}
\end{eqnarray}
where $\kappa_{+}= r_{+}^2\frac{\sqrt{f'(r_{+} h'(r_{+})}}{2}$ is the  surface gravity, the metric takes the following form,
\begin{equation}\label{}
	ds^2=-\mathcal{F}^2 dU dV +r^2(dx^2+dy^2).
\end{equation}
where the conformal factor is given by,
\begin{equation}\label{CFfun}
	\mathcal{F}^2=-\frac{4r^2 h(r)}{r_{+}^4f'(r_+)h'(r_+) U V}= r^2 h(r)e^{-2\kappa_{+} r^{*}(r)}
\end{equation}
The final metric form presents the original metric \eqref{metric1} written in the Kruskal-Szekeres coordinates.

	\bibliography{Island}

\end{document}